\documentclass[conference]{IEEEtran}
\IEEEoverridecommandlockouts
\usepackage{cite}
\usepackage{amsmath,amssymb,amsfonts}
\usepackage{algorithm}
\usepackage{graphicx}
\usepackage{subfigure}
\usepackage{textcomp}
\usepackage{xcolor}
\def\BibTeX{{\rm B\kern-.05em{\sc i\kern-.025em b}\kern-.08em
    T\kern-.1667em\lower.7ex\hbox{E}\kern-.125emX}}
\begin{document}

\title{A Novel JT-CoMP Scheme in 5G Fractal Small Cell Networks\\
\thanks{This work was supported in part by the National Key Research and Development Program of China under Grant 2017YFE0121600. Corresponding author: Xiaohu Ge.}
}

\author{\IEEEauthorblockN{1\textsuperscript{st} Jiaqi Chen}
\IEEEauthorblockA{\textit{School of Electronic Information and Communications} \\
\textit{Huazhong University of Science and Technology}\\
Wuhan, P. R. China \\
chenjq\_hust@hust.edu.cn}
\and
\IEEEauthorblockN{2\textsuperscript{nd} Xiaohu Ge}
\IEEEauthorblockA{\textit{School of Electronic Information and Communications} \\
\textit{Huazhong University of Science and Technology}\\
Wuhan, P. R. China \\
xhge@hust.edu.cn}
\and
\IEEEauthorblockN{3\textsuperscript{rd} Yi Zhong}
\IEEEauthorblockA{\textit{School of Electronic Information and Communications} \\
\textit{Huazhong University of Science and Technology}\\
Wuhan, P. R. China \\
yzhong@hust.edu.cn}
\and
\IEEEauthorblockN{4\textsuperscript{th} Yonghui Li}
\IEEEauthorblockA{\textit{School of Electrical and Information Engineering} \\
\textit{University of Sydney}\\
Sydney, Australia \\
yonghui.li@sydney.edu.au}
}

\maketitle

\begin{abstract}
To satisfy the requirement of the fifth generation (5G) mobile communications that offers an ultra high data rate of 100Mbps to 1Gbps anytime and anywhere, the coordinated multipoint (CoMP) technique is proposed to mitigate inter-cell interference to improve the coverage of high data rate services, cell-edge throughput, and system capacity. However, the joint transmission (JT) CoMP technique is difficult to be applied in practice due to the critical time synchronization for multiple coordination links and the bottleneck of backhaul capacity and radio resource at each small cell base stations (SBSs). Moreover, since the coordination SBSs in the conditional scheme are entirely separate from each other, different time of arrivals at the user cause the severe time synchronization problem. The anisotropic propagation environment in the urban scenario makes the implementation condition even worse. To tackle these issues, we propose a novel JT-CoMP scheme with the anisotropic path loss model to minimize the network backhaul traffic subject to the constraints on the radio resource and the differences in time of arrivals. Simulation results demonstrate that the proposed distance-resource-limited CoMP scheme can obtain the maximum achievable rate with the minimum network backhaul traffic, compared with existing schemes.
\end{abstract}


\section{Introduction}
The fifth generation (5G) mobile communication is designed to offer an experience data rate of 100Mbps to 1Gbps anytime and anywhere. The main problems are the severe channel conditions for the cell-edge users and the extremely high data rate requirements of the users in the central cell. The coordinated multipoint (CoMP) technique is believed to be an effective approach to mitigate inter-cell interference so as to improve the coverage of high data rate services, cell-edge throughput, and system capacity \cite{key-1,r1}. There are mainly three types of CoMP techniques: coordinated beamforming (CB), coordinated scheduling (CS), and joint transmission (JT). Among these CoMP techniques, the JT-CoMP technique offers the highest performance gains by fully sharing both data and control information among the coordinated base stations (BSs) \cite{key-2}. However, implementing JT-CoMP faces several critical challenges in practical 5G small cell networks. Firstly, JT-CoMP requires that the coordinated small cell base stations (SBSs) transmit data on the same time and frequency resources. There exist unavoidable differences in time of arrival (ToA) due to the different distances between the different transmitters and the receiver \cite{key-3,key-4}. Secondly, limited backhaul link creates a major bottleneck for high data information sharing among SBSs \cite{key-5,r2,r3}. Finally, for any SBS, the primary constraint is the limited radio resource. To fully explore the benefits of JT-CoMP, the above technical challenges must be carefully and effectively addressed.

There have been extensive studies on JT-CoMP in conventional cellular networks. A cooperation policy was introduced in \cite{key-6}, where the cooperation among BSs happens only when the user lies inside a planar zone at the cell borders. The decision for a user to choose service with or without cooperation was directed by a family of geometric policies, depending on its relative position to its two closest BSs. In \cite{key-7}, the three schemes to select coordinated SBSs in ultra-dense networks were compared. The scheme 1 selects SBSs with the highest value of the received power. In scheme 2, a user is associated with SBSs whose received power is above a given threshold. In scheme 3, a user is served by SBSs whose received powers are higher than a certain value, which is equal to the difference between the maximum received signal power and a threshold. The results indicated that these three JT-CoMP schemes had pros and cons in different aspects. An analytical model was proposed to perform adaptive modulation for a typical JT-CoMP system consisting of three transmission points \cite{key-2}. The closed-form expressions for the average spectral efficiency were obtained when adopting continuous-rate adaptive modulation. Authors in \cite{key-8} proposed a location-aware cross-tier BS cooperation scheme where the cooperation occurs among the macro cell and small cells and evaluated the performance regarding the outage probability and the average achievable rate.

Some optimizations in the JT-CoMP scenarios were carried out in the literature \cite{key-9,key-10,key-11,key-12}. To cope with the problem that the load-balancing capability becomes much lower than that expected in a clustered heterogeneous network (HetNet), a feasible suboptimal iterative algorithm was provided in \cite{key-9} for determining the joint user association solution of the self-organizing network. In \cite{key-10}, the authors optimized the coordinated cluster size and characterized the average downlink user data rate under a common non-coherent JT scheme, which was used to illustrate the trade-off between handoff rate and data rate. In air-to-ground cooperative communication
networks, an optimum altitude of the unmanned aerial vehicles for maximum coverage region was derived by guaranteeing a minimum outage performance over the region \cite{key-11}. The energy-spectral efficiency benefiting from the joint optimization of CoMP transmission and BS deployment was evaluated in the context of the dense large-scale cellular network \cite{key-12}.

However, the above studies all considered the isotropic propagation environments, where the path loss exponent is a constant on the whole plane. The actual propagation environment is very complicated, and a large number of line-of-sight (LoS) and non-line-of-sight (NLoS) transmissions exist. Our work in \cite{key-13} indicated that the
measured wireless cellular coverage boundary is extremely irregular with respect to directions, and the statistical fractal characteristic of coverage boundary exist widely, which confirmed that the actual propagation environment is anisotropic. Such a fractal characteristic is illustrated by the spectral density power-law behavior and the
slowly decaying variances in the angle domain \cite{key-21}. Moreover, in anisotropic propagation environments, the cooperation scheme with the received signal strength threshold can achieve higher user throughput rate than the cooperation scheme with the distance threshold when the number of coordinated SBSs is fixed to be equal in the two schemes \cite{key-14}. And the coordinated SBSs are distributed in different locations, which introduces different ToAs. In this case, the critical time synchronization is hard to satisfy in the anisotropic propagation environment. Therefore, how to select a coordinated set of SBSs in a complex propagation environment to meet the requirement of the time difference of arrivals, backhaul traffic limit, and radio resource limit, is a big challenge for JT-CoMP implementation in 5G fractal small cell networks. In this paper, we propose a novel scheme to minimize the network backhaul traffic with the limited radio resource constraint at SBS and subject to different distances between the coordinated SBSs and the served user, considering the anisotropic propagation environment. The main contributions of this work are summarized as follows.
\begin{itemize}
\item The three critical issues faced by the JT-CoMP technique are tackled under the anisotropic propagation environment in this paper, where the issues are the requirement of transmission delay difference, backhaul traffic limit, and radio resource limit.
\item A general JT-CoMP framework for minimizing the network backhaul traffic of 5G fractal small cell networks is formulated under the limited radio resource at SBS and different distances between the coordinated SBSs and the served user.
\item The distance-resource-limited CoMP scheme is proposed with the minimum network backhaul traffic compared with the baseline schemes in the simulation results.
\end{itemize}

The remainder of this paper is structured as follows. Section II describes the system model. The optimization for network backhaul traffic is investigated in Section III. The simulation results are discussed in Section IV. Finally, the conclusions are drawn in Section V.

\section{System Model}

\subsection{Network Model}

\begin{figure}[htbp]
\centerline{\includegraphics[width=8.5cm]{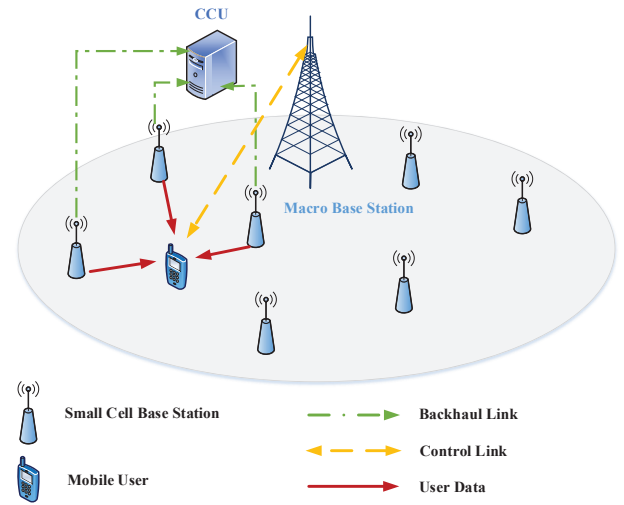}}
\caption{The CoMP transmission scenario based on CDSA.}
\label{fig:CDSA}
\end{figure}

We consider the downlink scenario of a two-tier HetNet in a control-data separation architecture (CDSA) \cite{key-15,key-16}, where small cells and users are covered by a macro base station (MBS), as illustrated in Fig. \ref{fig:CDSA}. The MBS is used to handle the control signaling and SBSs provide users with the required data services. Each MBS is equipped with a CoMP control unit (CCU) to provide smart clustering decisions centrally on the collaborative behavior of SBSs. The SBSs are connected to the CCU through wireless backhaul links and report their channel state information (CSI) to the CCU. The coordinated clustering among all SBSs is decided by the CCU based on the reported CSI.

Assume that the typical CDSA scenario consists of one MBS, $M$ SBSs and $K$ users. All SBSs and users are located randomly in the coverage area of the MBS. The locations of SBSs and users can be modeled as two independent Poisson point processes \cite{r4}, denoted as $\Phi_{B}$ and $\Phi_{U}$. The corresponding intensities of $\Phi_{B}$ and $\Phi_{U}$
are $\lambda_{B}$ and $\lambda_{U}$, respectively. In this paper, each user is assigned a cluster of SBSs by the CCU. The cluster of SBSs consists of the coordinated SBSs for the user. User data is available at all SBSs in the cooperation set through the backhaul links. The coordinated SBSs transmit the user data to the user at a time. The
cooperation set of the $k$-th user is denoted as $\Psi_{k}\subseteq\Phi_{B}$. We let the matrix $\mathbf{C}\in\mathrm{R}^{K\times M}$ denote the cooperative SBS clusters for all users, where the element $c_{km}\left(k=1,2,\cdots,K;\,m=1,2,\cdots,M\right)$ is an indicator function, having the value 1 when $\mathrm{SBS}_{m}$ belongs to $\Psi_{k}$ and the value 0 when $\mathrm{SBS}_{m}$ does not belong to $\Psi_{k}$, i.e.,
\begin{equation}
c_{km}=\begin{cases}
1 & \mathrm{SBS}_{m}\in\Psi_{k}\\
0 & \mathrm{SBS}_{m}\notin\Psi_{k}
\end{cases}.
\end{equation}
Furthermore, the number of the coordinated SBSs for the $k$-th user is denoted by $N_{k}^{B}={\sum_{m=1}^{M}c_{km}}$, and the number of the users simultaneously served by $\mathrm{SBS}_{m}$ is denoted by $N_{m}^{U}={\sum_{k=1}^{K}c_{km}}$.

\subsection{Anisotropic Path Loss Model}

The path loss in real environments is affected by electromagnetic radiation, the atmospheric environment, weather status, obstacle distribution, and diffraction and scattering effects. Researchers have examined the path loss exponent in some typical environments, e.g., an urban scenario with LoS transmissions, an urban scenario with NLoS
transmissions, and the near-ground propagation in the forest. The empirical path loss exponents for these three scenarios were reported as 2.1 \cite{key-22}, 3.19 \cite{key-22}, and 4 \cite{key-23}, respectively. Considering that obstacles consisting of buildings, plants, cars, human bodies and so on are irregularly distributed in
the real scenario, the links from a user to SBSs pass through different propagation environments. In this case, the path loss exponent of these links are different depending on the propagation environment. The path loss exponents of the different transmission links are assumed to be independently identically distributed (\textit{i.i.d.}) random
variables.

In the anisotropic scenario, the distances between the SBS and its coverage boundary was experimentally verified to have the fractal characteristic \cite{key-13}, and was modeled as a random variable with alpha-stable distribution \cite{key-21}. In \cite{key-24}, the path loss exponent of the link between a user and an SBS was derived to be the function of the distance between the SBS and its coverage boundary, which is expressed as $\beta=-{\lg\left(\frac{P_{min}}{P_{T}}\right)}/{\lg\left(R_{max}\right)}$, where $P_{min}$ is the minimum detectable signal power, $P_{T}$ is the transmit signal power of an SBS, $R_{max}$ is the distance between the SBS and its coverage boundary. Assume that the minimum detectable signal powers of all users are assumed to be equal and all SBSs transmit signals with the same power. In this case, $\lg\left(\frac{P_{min}}{P_{T}}\right)$ is a constant denoted by $\zeta$. The probability density function (PDF) of $R_{max}$ is expressed in \cite{key-24} as
\begin{equation}
f\left(R_{max}\right)=\begin{cases}
\frac{\varepsilon}{\rho_{min}^{-\varepsilon}-\rho_{max}^{-\varepsilon}}R_{max}^{-\left(\varepsilon+1\right)}, & \rho_{min}\leq R_{max}\leq\rho_{max}\\
0, & \mathrm{otherwise}
\end{cases},\label{eq:R_PDF}
\end{equation}
where $\varepsilon\in\left(1,2\right]$ is the fractal parameter. Based on (\ref{eq:R_PDF}), the PDF of the path loss exponent can be derived as
\begin{equation}
f\left(\beta\right)=\begin{cases}
\frac{-\varepsilon\ln10}{\rho_{min}^{-\varepsilon}-\rho_{max}^{-\varepsilon}}10^{\frac{\zeta}{\beta}\varepsilon}\frac{\zeta}{\beta^{2}}, & \frac{-\zeta}{\lg\left(\rho_{max}\right)}\leq\beta\leq\frac{-\zeta}{\lg\left(\rho_{min}\right)}\\
0, & \mathrm{otherwise}
\end{cases}.\label{eq:PLE_PDF}
\end{equation}

Furthermore, the received signal power at the $k$-th user from $\mathrm{SBS}_{m}$ is expressed as
\begin{equation}
P_{km}=P_{T}h_{km}r_{km}^{-\beta_{km}},\label{eq:recceived signal power}
\end{equation}
where $h_{km}$ denotes the power gain from Rayleigh fading between $\mathrm{SBS}_{m}$ and the $k$-th user, $r_{km}$ denotes the distance between $\mathrm{SBS}_{m}$ and the $k$-th user, $\beta_{km}$ is the path loss exponent. Considering that the JT-CoMP technique is adopted, the received signal power of the $k$-th user is given as
\begin{align}
P_{k} & =\underbrace{{\displaystyle \sum_{\mathrm{SBS}_{m}\in\Psi_{k}}P_{km}}}+\underbrace{{\displaystyle \sum_{\mathrm{SBS}_{n}\in\Phi_{B}-\Psi_{k}}P_{kn}}}+\sigma_{n}^{2},\\
 & \mathrm{\quad Desired\,signal}\qquad\mathrm{Interference}\nonumber
\end{align}
where $\Phi_{B}-\Psi_{k}$ represents the cochannel interfering SBSs outside the cooperation set $\Psi_{k}$, $\sigma_{n}^{2}$ is the noise power at the user. The received SINR at the $k$-th user is expressed as
\begin{align}
\mathrm{SINR}_{k} & =\frac{\sum_{\mathrm{SBS}_{m}\in\Psi_{k}}P_{km}}{\sum_{\mathrm{SBS}_{n}\in\Phi_{B}-\Psi_{k}}P_{kn}+\sigma_{n}^{2}}\nonumber \\
 & =\frac{\sum_{\mathrm{SBS}_{m}\in\Phi_{B}}c_{km}P_{km}}{\sum_{\mathrm{SBS}_{n}\in\Phi_{B}}P_{kn}-\sum_{\mathrm{SBS}_{m}\in\Phi_{B}}c_{km}P_{km}+\sigma_{n}^{2}}.\label{eq:SINR}
\end{align}
All SBSs are assumed to share the same physical resource blocks (PRB). Each PRB can be assigned to only one user for simplicity. The achievable rate at the $k$-th user over a PRB with the bandwidth $B$ is
\begin{equation}
R_{k}=B\log_{2}\left(1+\mathrm{SINR}_{k}\right).\label{eq:rate}
\end{equation}

\section{Optimization for Network Backhaul Traffic}

\subsection{Network Backhaul Traffic}

In order to improve user data rate and spectrum efficiency, more SBSs are assigned to serve the user cooperatively. However, when more SBSs share CSI and user data, more traffic will be generated in the backhaul links, which generates the requirements on high capacity of the backhaul links. The required backhaul capacity for an SBS is determined by two factors: the number of users associated with an SBS, and the amount of data need to be transmitted through the backhaul network for a user. The data shared among coordinated SBSs can be divided into two parts, i.e., CSI and user data. The amount of CSI is negligible compared to the amount of user data \cite{key-7}. The backhaul traffic of $\mathrm{SBS}_{m}$ is expressed as
\begin{equation}
R_{m}^{b}={\sum_{k=1}^{K}c_{km}\gamma_{k}R_{k}},\label{eq:bh}
\end{equation}
where $\gamma_{k}$ is the required number of PRBs of the $k$-th user. In this paper, the small cell network is configured to satisfy the quality of service (QoS) of users when the achievable rates of all users are larger than the minimum traffic requirements. The minimum traffic requirement of a user is denoted as $R^{min}$. The number of PRBs assigned to the $k$-th user should be
\begin{equation}
\gamma_{k}=\left\lceil \frac{R^{min}}{R_{k}}\right\rceil ,
\end{equation}
where $\left\lceil \cdot\right\rceil $ is the ceiling operation. The coordinated SBSs transmit data to a user on the same time and frequency resources. In this case, the same $\gamma_{k}$ PRBs of each SBS in the cooperation set $\Psi_{k}$ are assigned to the $k$-th user. The network backhaul traffic is given as the average backhaul traffic of each SBS in the small cell network, denoted as $R^{b}=\mathbb{E}_{\Phi_{B}}\left[R_{m}^{b}\right]$.

\subsection{Selection for Cooperation Set}

Since the available radio resource at an SBS is limited that cannot supply every user with the enough resource. When an SBS works at its full capacity, other user can't associate itself with the SBS, which results in that the user can't achieve the desired rate \cite{r5}. It is of a major significance to propose a novel JT-CoMP scheme with the limited
radio resource at the SBS\cite{r6,r7}. As such, a constraint on the number of PRBs is considered, where the required number of PRBs at $\mathrm{SBS}_{m}$ is given as $N_{m}^{\Omega}=\sum_{k=1}^{K}c_{km}\gamma_{k}$, with $c_{km}\in\{0,1\}$. Assumed that the total number of PRBs provided by each SBS is $\Omega$. We have a constraint that $N_{m}^{\Omega}\leq\Omega$. What's more, a user is assigned at least one PRB, \textit{i.e.}, $\gamma_{k}\geq1,$ and $N_{k}^{B}\geq1$, $\forall k\in\left\{ 1,2,\ldots,K\right\} $.

Due to the inter-symbol interference caused by the existence of multipath delays, the system cyclic prefix (CP) is proposed to control the negative impact of multipath delay. When the multipath delay is smaller than the CP, the receiver can catch the full energy of the orthogonal frequency division multiplexing symbol. When the multipath delay is larger than CP, the energy of some symbols cannot be caught by the receiver, and even the energy of the previous symbol is received by the receiver as the next symbol which results in the inter-symbol interference. In JT-CoMP, multiple SBSs provide a user with shared data at the same time. For users, this is equivalent to a multiple-transmitting-single-receiving system. Considering that the geographic locations of the coordinated SBSs are separated from each other, the distances between coordinated SBSs and the user are different. Therefore, the delay between each coordinated SBS and the user must meet the requirement of the system CP, \textit{i.e.}, the difference in ToAs should be less than the system's CP. If the difference in ToAs is greater than the system CP, JT-CoMP will fail. The Long Term Evolution (LTE) system has determined that the CP length of a regular cell is 4.6875$\mathrm{\mu s}$, but in 5G communication systems, this requirement will be more critical, up to about 500$\mathrm{ns}$ \cite{key-4}. In this paper, the difference in ToAs among the coordinated SBSs is equivalent to the difference in the distances between coordinated SBSs and the user, expressed as $\left|r_{km}-r_{kn}\right|\leq ct_{cp},\,\forall\mathrm{SBS}_{m},\mathrm{SBS}_{n}\in\Psi_{k}$, where $c$ is the light speed equal to $3\times10^{8}\mathrm{m/s}$ and $t_{cp}$ is the CP length equal to 500$\mathrm{ns}$.

The problem of interest is to minimize the network backhaul traffic of a fractal small cell network with the JT-CoMP technique, which is essentially a weighted sum-rate of users presented in the network. Mathematically, the network backhaul traffic minimization problem can be expressed as
\begin{align}
 & \underset{\mathbf{C}}{\textrm{min}}R^{b},\label{eq:object}\\
 s.t. \quad & \gamma_{k}R_{k}\geq R^{min},\,\forall k\in\left\{ 1,2,\ldots,K\right\} ,\\
 & 1\leq\sum_{k=1}^{K}c_{km}\gamma_{k}\leq\Omega,\,\forall\mathrm{SBS}_{m}\in\Phi_{B},\\
 & \gamma_{k}\geq1,\,\forall k\in\left\{ 1,2,\ldots,K\right\} ,\\
 & \sum_{m=1}^{M}c_{km}\geq1,\,\forall k\in\left\{ 1,2,\ldots,K\right\} ,\\
 & \left|r_{km}-r_{kn}\right|\leq ct_{cp},\,\forall\mathrm{SBS}_{m},\mathrm{SBS}_{n}\in\Psi_{k}.
\end{align}

\subsection{Distance-Resource-Limited CoMP Scheme}

As shown in (\ref{eq:bh}), in order to effectively reduce the network backhaul traffic that is a weighted sum of the numbers of coordinated SBSs and corresponding achievable rates, generally the SBS with the strongest received signal power should be assigned to the user. What's more, the achievable rate at the user is given as (\ref{eq:rate})
with (\ref{eq:SINR}). It can be found that the backhaul traffic is not a linear function with respect to the coordination matrix $\mathbf{C}$. The theoretical optimization result is hard obtained.

In this section, we propose a scheme considering the limits of the distance and resource of the SBSs, named distance-resource-limited CoMP scheme, to find a cooperation set for each user such that the network backhaul traffic is minimized. The proposed scheme is designed as the two steps. 1) The prior user should be selected for servicing by multiple SBSs. Because the radio resource blocks are limited, the more SBSs should serve the users whose channel conditions are bad such that the users with better channel conditions are served by only one SBS to save the PRBs of the whole network. A coordination priority of users is configured as
\begin{equation}
\Lambda_{k}=\frac{R^{min}}{\log_{2}\left(1+\mathrm{SINR}_{k}^{0}\right)},
\end{equation}
where $\mathrm{SINR}_{k}^{0}$ is the SINR of the $k$-th user without cooperation considering that the user associates itself with the SBS $\mathrm{SBS}_{0}$ providing the strongest signal power, expressed as
\begin{equation}
\mathrm{SINR}_{k}^{0}=\frac{P_{T}h_{k0}r_{k0}^{-\beta_{k0}}}{\sum_{m=1}^{M}P_{T}h_{km}r_{km}^{-\beta_{km}}-P_{T}h_{k0}r_{k0}^{-\beta_{k0}}+\sigma_{n}^{2}}.\label{eq:SINR0}
\end{equation}
The larger value of $\Lambda_{k}$ denotes the higher priority, i.e., the user with larger value of $\Lambda_{k}$ should be considered first. 2) Multiple SBSs should be selected to form a cooperation set and serve a user. The SBSs with the stronger received signal powers should be assigned to serve the user. And the distances between any
two coordinated SBSs are limited. The remained PRBs at all SBSs should be considered.

The pseudo code of the proposed distance-resource-limited CoMP scheme is described in Algorithm 1. Firstly, the priorities $\Lambda_{k}\left(k=1,2,\ldots,K\right)$ for all users are calculated (Line 1--4). The user index $k$ is sorted in descending order of the priority $\Lambda_{k}$ (Line 5). In other words, if the user with a larger value of $\Lambda_{k}$, we will firstly find a JT cooperation set for serving the user. Then, the cooperation sets for all users are found. For the $k$-th user, the SBS with the strongest received signal power is selected for cooperation (Line 6--11), and the required PRBs is calculated. If the distances between the particular SBS and the SBSs in the cooperation set are smaller than $ct_{cp}$ (Line 12--14), and the required PRBs are smaller than the remained PRBs of the particular SBS (Line 16\textendash 25), this SBS is added into the cooperation set. In line 12, the distance constraint between any two coordinated SBSs is set as $ct_{cp}/2$ to ensure that the maximum distance among the coordinated SBSs is smaller than $ct_{cp}$. When the while loop is finished, the minimum network backhaul traffic of the fractal small cell network is satisfied.

\begin{algorithm}[htbp]
\caption{Distance-resource-limited\textbf{ }CoMP scheme. }

\textbf{Input:} $\Phi_{B}$, $\Phi_{U}$, $\mathbf{H}=\left[h_{km}\right]_{K\times M}$,$\mathbf{A}=\left[\beta_{km}\right]_{K\times M}$,
$\Omega$

\textbf{Output:} $\mathbf{C}=\left[c_{km}\right]_{K\times M}$

1: \textbf{for} $k=1:K$ \textbf{do}

2:Calculate (\ref{eq:SINR0});

3: $\Lambda_{k}=\frac{R^{min}}{\log_{2}\left(1+\mathrm{SINR}_{k}^{0}\right)}$;

4: \textbf{end}

5: Sort user index $k$ in descending order of priority $\Lambda_{k}$;

6: $\Psi_{k}\leftarrow\varnothing,\mathbf{C}\leftarrow\left[0\right]_{K\times M}$;

7: $\mathbf{P}=\left[P_{km}\right]_{K\times M}$, $P_{km}\leftarrow P_{T}h_{km}r_{km}^{-\beta_{km}}$;

8: \textbf{for} $k=1:K$ \textbf{do}

9: $T\leftarrow1$;

10: \textbf{while} $\sum_{m=1}^{M}c_{km}<T$ \textbf{do}

11: find the index $m$ with the $T$-th largest
$P_{km}$

12: \textbf{if} $\left|r_{km}-r_{k0}\right|>ct_{cp}/2$
\textbf{then}

13: $T\leftarrow T+1$;

14: Go to Line 10;

15: \textbf{else}

16: calculate the achievable rate $R_{k}$
with $\Psi_{k}$;

17:  calculate the required PRB $\gamma_{k}$;

18:  \textbf{if} $\gamma_{k}$ is smaller
than the remained PRB of $\mathrm{SBS}_{m}$ \textbf{then}

19:  $\Psi_{k}\leftarrow\Psi_{k}\cup\mathrm{SBS}_{m}$;

20: $c_{km}\leftarrow1;$

21:\textbf{if} $\gamma_{k}R_{k}>R^{min}$
\textbf{then}

22:\textbf{end while}

23: \textbf{else}

24:  $T\leftarrow T+1$;

25:  Go to Line 10;

26: \textbf{end for}

27: \textbf{return} $\mathbf{C}$
\end{algorithm}

The time complexity of Algorithm 1 is $O\left(KM^2\right)$. For each user, Algorithm 1 executes the coordinated SBS selection for a user in an iteration (Line 12-25). In the selection, we take $O\left(M\right)$ time to find the coordinated SBSs. Since the while loop will be executed at most $M$ times, the function can be done in $M^2$ times (Lines 11). Therefore, Algorithm 1 takes $O\left(M^2\right)$ to find a joint transmission cluster for a user. Since there are at most $K$ users, the time complexity of Algorithm 1 is $O\left(KM^2\right)$. The time complexity of the best power cooperation scheme and that of the best distance cooperation scheme are $O\left(KM\right)$.

\section{Simulation Results}

In this section, the proposed scheme is analyzed and compared by simulation results. To better justify our motivation and observations, the proposed distance-resource-limited CoMP scheme, denoted by ``DRC'' in figures, is compared with the best power cooperation scheme (denoted by BPC scheme) where the user is associated with SBSs whose received signal power is larger than the threshold $T$, and the best distance cooperation scheme (denoted by BDC scheme) where the user is associated with the $K$ nearest SBSs. The default parameters are configured as the Table I \cite{key-14,key-15}.
\begin{table}[htbp]
\caption{PARAMETER SETTINGS}
\centering{}%
\begin{tabular}{|l|l|}
\hline
\textbf{Parameter} & \textbf{Value}\tabularnewline
\hline
\hline
The density of SBSs $\lambda_{B}$ & $10^{-4}$/$\mathrm{m^{2}}$\tabularnewline
\hline
The density of users $\lambda_{U}$ & $10^{-3}$/$\mathrm{m^{2}}$\tabularnewline
\hline
The transmit power $P_{T}$ & 1W\tabularnewline
\hline
The number of PRBs at an SBS $\Omega$ & 25\tabularnewline
\hline
The noise power $\sigma_{n}^{2}$ & -95dBm\tabularnewline
\hline
The minimum traffic $R^{min}$ & 4bits/s/Hz\tabularnewline
\hline
The CP length $t_{cp}$ & 500$\mathrm{ns}$\tabularnewline
\hline
The power threshold $T$  & -70dBm\tabularnewline
\hline
The number of coordinated SBSs $K$ & 3\tabularnewline
\hline
\end{tabular}
\end{table}

%
%

\begin{figure} 
\centering 
\subfigure[]{ 
\includegraphics[width=8.5cm]{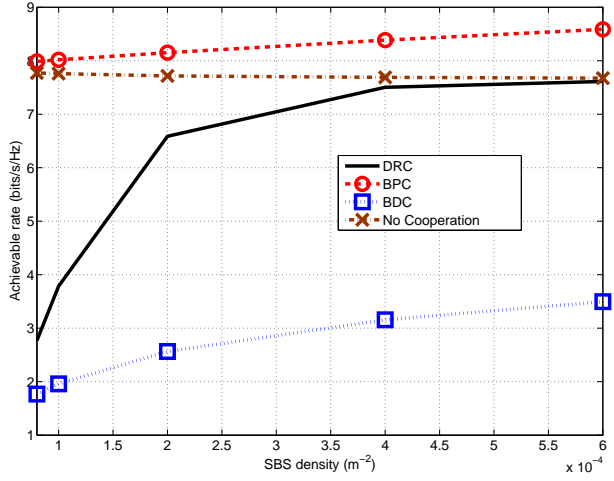}} 
\hspace{1in} 
\subfigure[]{ 
\includegraphics[width=8.5cm]{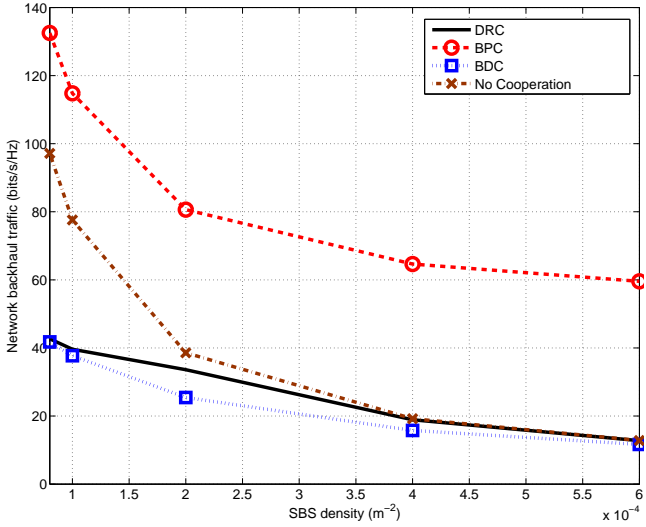}} 
\caption{The achievable rate and network backhaul traffic with respect to the
SBS density based on the distance-resource-limited CoMP scheme, compared
with the BPC and BDC schemes.}  \label{fig:AR-lambda} 
\end{figure}
Fig. \ref{fig:AR-lambda} shows the achievable rate and network backhaul
traffic with respect to the SBS density based on the distance-resource-limited
CoMP scheme. And the results are compared with that of the BPC and
BDC schemes. In Fig. 2(a), the achievable rate increases with the
increase of the SBS density. It is found that the achievable rate
of the BDC scheme is extremely lower than that of the proposed scheme.
Because some of SBSs nearest the user may be NLoS SBSs and can not
provide the strongest received signal power in the anisotropic propagation
environment. In extreme cases, the desired signal power is smaller
than the interfering signal power for the BDC scheme. In this case,
the proposed scheme and BPC scheme can ensure that the coordination
SBSs provide the strongest signal power. What's more, the achievable
rate of the BPC scheme is better than the proposed scheme. In Fig.
2(b), the network backhaul traffic decreases with the increase of
the SBS density. The network backhaul traffic of the BPC scheme is
much larger than that of the proposed scheme. The results in Fig.
\ref{fig:AR-lambda} imply that the high achievable rate can be obtained
based on the distance-resource-limited CoMP scheme while causing low
network backhaul traffic.

\begin{figure}
\centering
\subfigure[]{
\includegraphics[width=8.5cm]{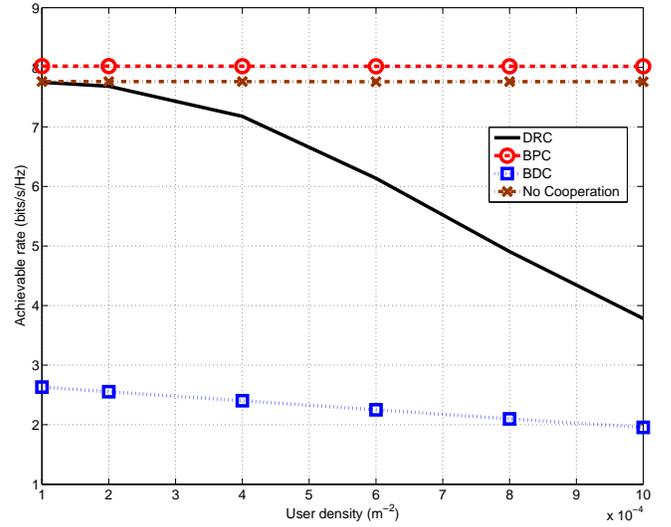}}
\hspace{1in}
\subfigure[]{
\includegraphics[width=8.5cm]{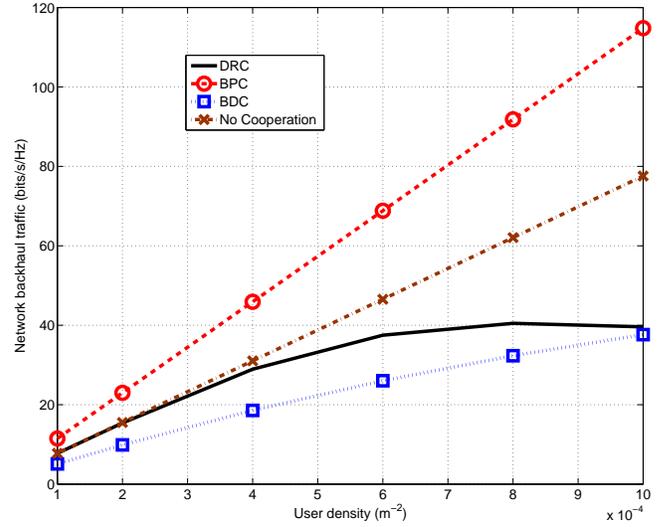}}
\caption{The achievable rate and network backhaul traffic with respect to the user density based on the distance-resource-limited CoMP scheme, compared with the BPC and BDC schemes.}  \label{fig:NB-U} 
\end{figure}
Fig. \ref{fig:NB-U} shows the achievable rate and network backhaul
traffic with respect to the user density based on the distance-resource-limited
CoMP scheme, compared with the BPC and DC schemes. In Fig. 3(a), the
achievable rate decreases with the increase of the user density. It
is because that the number of PRBs assigned to each user is not enough
to obtain the required data rate with the number of user covered by
an MBS increasing. In Fig. 3(b), the network backhaul traffic increases
with the increase of the user density. When the user density is equal
to the SBS density, i.e. $\lambda_{U}=\lambda_{B}=10^{-4}\mathrm{/m^{2}}$,
the network backhaul traffics of the three schemes are same while
the achievable rates of the distance-resource-limited CoMP scheme
and the BPC scheme are higher than that of the BDC scheme.

\begin{figure}[tbh]
\begin{centering}
\includegraphics[width=8cm]{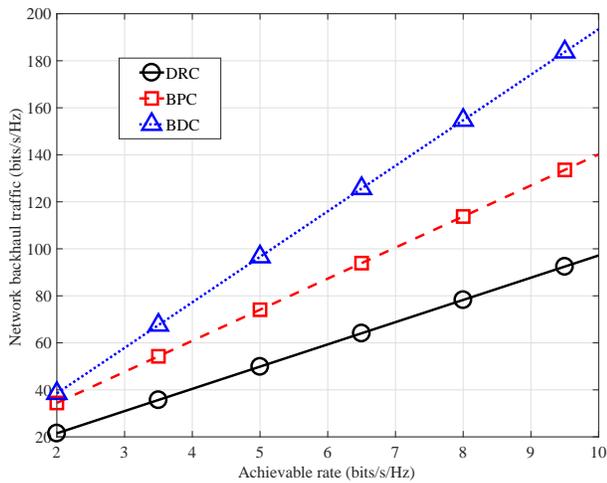}
\par\end{centering}
\caption{The network backhaul traffic with respect to the achievable rate based
on the distance-resource-limited CoMP scheme, compared with the BPC
and BDC schemes.\label{fig:ntBH_AchR}}
\end{figure}
Fig. \ref{fig:ntBH_AchR} shows the network backhaul traffic with
respect to the achievable rate based on the distance-resource-limited
CoMP scheme, compared with the BPC and BDC schemes. The network backhaul
traffic increases with the increase of the achievable rate. When the
achievable rate is fixed, the network backhaul traffic of the proposed
scheme is smaller than that of the other two schemes. When the network
backhaul traffic is fixed, the achievable rate of the proposed scheme
is higher than that of the other two schemes. The result implies that
the proposed scheme costs the minimum network backhaul traffic to
obtain the same achievable rate compared with the BPC and BDC schemes,
which meets the original intention of this paper.

\section{Conclusions}
To satisfy the requirement of the 5G mobile communication that offers
a data rate of 100Mbps to 1Gbps anytime and anywhere, the CoMP technology
is proposed to mitigate inter-cell interference which results in improving
the coverage of high data rate services, cell-edge throughput, and
system capacity. In this paper, the JT-CoMP technique is targeted
considering the difference in time of arrivals for multiple coordination
links and the limited radio resource at each SBSs. Firstly, a general
JT-CoMP framework for minimizing the network backhaul traffic of 5G
fractal small cell networks is formulated subjected to a limited radio
resource at SBS and the difference in distances between the coordinated
SBSs and the served user. Then, we propose a distance-resource-limited
CoMP scheme considering the anisotropic path loss model to minimize
the network backhaul traffic. Simulation results show that the proposed
distance-resource-limited CoMP scheme obtains the high achievable
rate with the minimum network backhaul traffic, compared with the
two existing schemes.


\begin{thebibliography}{00}

\bibitem{key-1}X. Ge, S. Tu, G. Mao, C.-X. Wang and T. Han, ``5G Ultra-Dense Cellular Networks,¡± IEEE Wireless Commun., vol. 23, no. 1, pp.72--79, Feb. 2016.
\bibitem{r1}X. Ge, J. Ye, Y. Yang and Q. Li, ``User Mobility Evaluation for 5G Small Cell Networks Based on Individual Mobility Model," IEEE J. Sel. Areas Commun., vol. 34, no. 3, pp. 528--541, Mar. 2016.
\bibitem{key-2}Q. Cui, H. Song, H. Wang, M. Valkama and A. A. Dowhuszko, ``Capacity Analysis of Joint Transmission CoMP With Adaptive Modulation," IEEE Trans. Veh. Technol., vol. 66, no. 2, pp. 1876--1881, Feb. 2017.
\bibitem{key-3}V. Kotzsch and G. Fettweis, ``On synchronization requirements and performance limitations for CoMP systems in large cells," in 8th International Workshop on Multi-Carrier Systems \& Solutions, Herrsching, Germany, May 2011.
\bibitem{key-4}D. P. Venmani and A. Kaoru, ``On Improving Clock Synchronization Accuracy for LTE-A Networks," in IEEE 82nd Vehicular Technology Conference, Boston, MA, USA, Sept. 2015.
\bibitem{key-5}D. Lee et al., ``Coordinated multipoint transmission and reception in LTE-advanced: deployment scenarios and operational challenges," IEEE Commun. Mag., vol. 50, no. 2, pp. 148--155, Feb. 2012.
\bibitem{r2}X. Ge, H. Cheng, M. Guizani, T. Han, ``5G Wireless Backhaul Networks: Challenges and Research Advances," IEEE Network, vol. 28, no. 6, pp. 6--11, Nov. 2014.
\bibitem{r3}X. Ge, S. Tu, T. Han, Q. Li and G. Mao, ``Energy Efficiency of Small Cell Backhaul Networks Based on Gauss-Markov Mobile Models," IET Networks, vol. 4, no. 2,  pp. 158--167, 2015.
\bibitem{key-6}F. Baccelli and A. Giovanidis, ``A Stochastic Geometry Framework for Analyzing Pairwise-Cooperative Cellular Networks," IEEE Trans. Wireless Commun., vol. 14, no. 2, pp. 794--808, Feb. 2015.
\bibitem{key-7}S. Chen, T. Zhao, H. H. Chen, Z. Lu and W. Meng, ``Performance Analysis of Downlink Coordinated Multipoint Joint Transmission in Ultra-Dense Networks," IEEE Network, vol. 31, no. 5, pp. 106--114, Aug. 2017.
\bibitem{key-8}A. H. Sakr and E. Hossain, ``Location-Aware Cross-Tier Coordinated Multipoint Transmission in Two-Tier Cellular Networks," IEEE Trans. Wireless Commun., vol. 13, no. 11, pp. 6311--6325, Nov. 2014.
\bibitem{key-9}J. B. Park and K. S. Kim, ``Load-Balancing Scheme With Small-Cell Cooperation for Clustered Heterogeneous Cellular Networks," IEEE Trans. Veh. Technol., vol. 67, no. 1, pp. 633--649, Jan. 2018.
\bibitem{key-10}W. Bao and B. Liang, ``Optimizing Cluster Size Through Handoff Analysis in User-Centric Cooperative Wireless Networks," IEEE Trans. Wireless Commun., vol. 17, no. 2, pp. 766--778, Feb. 2018.
\bibitem{key-11}M. M. Azari, F. Rosas, K. C. Chen and S. Pollin, ``Ultra Reliable UAV Communication Using Altitude and Cooperation Diversity," IEEE Trans. Commun., vol. 66, no. 1, pp. 330--344, Jan. 2018.
\bibitem{key-12}G. Zhao, S. Chen, L. Zhao and L. Hanzo, ``Joint Energy-Spectral-Efficiency Optimization of CoMP and BS Deployment in Dense Large-Scale Cellular Networks," IEEE Trans. Wireless Commun., vol. 16, no. 7, pp. 4832--4847, Jul. 2017.
\bibitem{key-13}X. Ge et al., ``Wireless fractal cellular networks" IEEE Wireless Commun., vol. 23, no. 5, pp. 110--119, Oct. 2016.
\bibitem{key-21}X. Ge et al., ``Wireless Single Cellular Coverage Boundary Models" IEEE Access, vol. 4, pp. 3569--3577, Jul. 2016.
\bibitem{key-14}F. Bin, J. Chen, X. Ge and W. Xiang, ``Downlink Small-Cell Base Station Cooperation Strategy in Fractal Small-Cell Networks," in IEEE GLOBECOM, Singapore, Singapore, Dec. 2017.
\bibitem{key-15}S. Bassoy, M. Jaber, M. A. Imran and P. Xiao, ``Load Aware Self-Organising User-Centric Dynamic CoMP Clustering for 5G Networks," IEEE Access, vol. 4, pp. 2895--2906, 2016.
\bibitem{key-16}A. Mohamed, O. Onireti, M. A. Imran, A. Imran and R. Tafazolli, ``Control-Data Separation Architecture for Cellular Radio Access Networks: A Survey and Outlook," IEEE Commun. Surveys \& Tutorials, vol. 18, no. 1, pp. 446--465, First quarter 2016.
\bibitem{r4}X. Ge, B. Yang, J. Ye, G. Mao, C.-X. Wang and T. Han, ``Spatial Spectrum and Energy Efficiency of Random Cellular Networks," IEEE Trans. Commun., vol. 63, no. 3, pp. 1019--1030, Mar. 2015.
\bibitem{key-22}3GPP TR 38.900, ``Technical Specification Group Radio Access Network; Study on channel model for frequency spectrum above 6 GHz," Jul. 2017.
\bibitem{key-23}Y. S. Meng, Y. H. Lee and B. C. Ng, ``Empirical Near Ground Path Loss Modeling in a Forest at VHF and UHF Bands," IEEE Trans. Antennas Propag., vol. 57, no. 5, pp. 1461--1468, May 2009.
\bibitem{key-24} X. Ge, X. Tian, Y. Qiu, G. Mao and T. Han, ``Small Cell Networks with Fractal Coverage Characteristics," IEEE Trans. Commun., vol. 66, no. 11, pp. 5457--5469, Nov. 2018.
\bibitem{r5} X. Ge et al., ``Energy Efficiency Optimization for MIMO-OFDM Mobile Multimedia Communication Systems with QoS Constraints," IEEE Trans. Veh. Technol., vol. 63, no. 5, pp. 2127--2138, Jun. 2014.
\bibitem{r6}L. Xiang, X. Ge, C.-X. Wang, Frank Y. Li and Frank Reichert, ``Energy Efficiency Evaluation of Cellular Networks Based on Spatial Distributions of Traffic Load and Power Consumption," IEEE Trans. Wireless Commun., vol. 12, no. 3, pp.961--973, Mar. 2013.
\bibitem{r7}X. Ge, K. Huang, C.-X. Wang, X. Hong, and X. Yang, ``Capacity analysis of a multi-cell multi-antenna cooperative cellular network with co-channel interference," IEEE Trans. Wireless Commun., vol. 10, no. 10, pp.3298--3309, Oct. 2011.

\end{thebibliography}
\end{document}